\documentstyle[aps,epsf,prd]{revtex}
\begin{document}
\begin{flushright} {OITS 645}\\
August 1998
\end{flushright}
\vspace*{1cm}

\begin{center}  {\large {\bf A Color Mutation Model of Soft
Interaction\\ in High Energy Hadronic Collisions}}
\vskip .75cm
 {\bf  Zhen Cao{\footnote{Present address: Department of Physics,
University of Utah, Salt Lake City, UT 84112.}} and Rudolph C.
Hwa{\footnote{E-mail address: hwa@oregon.uoregon.edu}}}
\vskip.5cm
 {\bf Institute of Theoretical Science and Department of Physics\\
University of Oregon, Eugene, OR 97403-5203, USA}
\end{center}

\begin{abstract}
A comprehensive model, called ECOMB, is proposed to describe
multiparticle production by soft interaction. It incorporates the eikonal
formalism, parton model, color mutation, branching and recombination.
The physics is conceptually opposite to the dynamics that underlies the
fragmentation of a string. The partons are present initially in a hadronic
collision; they form a single, large, color-neutral cluster until color
mutation of the quarks leads to a fission of the cluster into two
color-neutral subclusters. The mutation and branching processes
continue until only $q\bar q$ pairs are left in each small cluster. The
model contains self-similar dynamics and exhibits scaling behavior in the
factorial moments. It can satisfactorily reproduce the intermittency data
that no other model has been able to fit.

\end{abstract}

\section{Introduction}

The study of multiparticle production in low-$p_T$ processes has been
pursued for over twenty years.  Since they involve soft interactions,
they cannot be treated in perturbative QCD.  In the absence of any
reliable theoretical approach to the problem, many models have been
proposed, most of which are represented in the review volumes
published ten years ago \cite{car,hx}.  Nearly all of those models have
since been shown to be inadequate in light of the data on fluctuations
and intermittency \cite{ddk,see}.  Indeed, there are very few models
that have the appropriate dynamical content capable of reproducing
the scaling behaviors observed in the experiments.  Even though
perturbative QCD cannot be used, it seems that some kind of branching
process is needed to generate the property of self-similarity, as the
resolution scale is varied.  In this paper we propose a model that
incorporates some aspects of nonperturbative QCD and is capable of
generating the features of the intermittency data, which are shown in
the last figure of this paper.  To our knowledge those data have not
been fitted by any model that contains some features of the color
dynamics.  To reproduce those data has become the main motivator
for this work.

Our approach embraces many time-honored properties of hadronic
collisions.  Since hadrons are extended objects, the eikonal formalism is
at the foundation of our model.  Thus it is not difficult for our model to
possess the virtues of geometrical scaling \cite{ddd} and approximate KNO
scaling \cite{kn}.  In order to build into the model features of
chromodynamics, it is necessary to introduce quarks and gluons into
the eikonal formalism, so the parton model is an essential gateway
into the microscopic domain of color interactions.  Once we enter that
domain, we embark on an unconventional journey of studying color
mutation of the constituents as a dynamical process by which the
colors of the quarks evolve through the emission and absorption of
gluons.  The smallness of
$\alpha_S$ is never assumed, so the evolution is not perturbative.
With all partons taken into consideration globally, we follow the
evolution of the configuration in the color space.  When the
configuration exhibits color neutral subclusters, we allow branching to
take place with a possible contraction  of the cluster size in
accordance to a reasonable rule consistent with confinement
dynamics. Successive branching leads to smaller and smaller clusters,
until they are finally identified as particles and resonances. The decay of
resonances are also taken into account before the final state of an
event is determined. Evidently, the model contains many features of
soft interaction that are familiar and desirable at a qualitative level.
Here we put them on a quantitative basis.

We shall call this model ECOMB, which stands for eikonal color
mutation branching. Only the eikonal part overlaps with an earlier
model, called ECCO, an eikonal cascade code \cite{hp}. Whereas
branching is put in by hand in ECCO, it is a consequence of the color
dynamics in ECOMB. Since the partons play a fundamental role in this
model, it is significantly closer to QCD than any eikonal model on soft
hadronic collisions has ever been.

There are a few parameters in the model. They are adjusted to fit  a
large body of experimental data on low-$p_T$ processes for
$\sqrt{s} \stackrel{<}{\sim} 100$ GeV. They include
$\sigma_{el}$,
$\sigma_{inel}$,
$\left<n\right>$, $C_q$, $dn/dy$, $P_n$, and $F_q$ for all $s$ in the
range
$10 < \sqrt{s} <100$ GeV, and all rapidity intervals. $F_q$ are the
normalized factorial moments, whose power-law dependence on the
rapidity bin-size $\delta$ has been referred to as the intermittency
behavior \cite{bp}.  Except for $F_q$, all the other pieces of datga are
global in nature; i.e., examined in or averaged over all rapidity space.
They can be fitted by many models.  $F_q$ in small rapidity intervals
exhibit local fluctuations, which are what invalidate most of those
models.

There is one important aspect about our model that should be
commented on in these introductory remarks. ECOMB is built upon the
parton model, and therefore represents an approach to multiparticle
production at low $p_T$ that is opposite to that of the more familiar
models, such as the Lund string model \cite{lun}, and the dual parton
model \cite{dpm}. The basic idea of those models is that the
momentum of an incident hadron is carried mainly by a few of the
valence quarks, which upon collision drag a color flux tube or a string
that subsequently fragment due to confinement forces. The parton
model, on the other hand, as originally proposed by Feynman for soft
processes \cite{fey}, regards the hadron momentum as being carried
by all partons, whose momentum distribution is frozen by a hard
collision  process (except for $Q^2$ evolution). For production at low
$p_T$ the partons cannot be regarded as being momentarily free. It
does not mean that the parton model is invalid; it only means that the
model cannot be naively applied, as in hard processes. In describing
the inclusive cross section of  $pp$ collision in the large $x_F$ region,
the Lund model \cite{lun} had one extreme view in that the produced
hadrons are the result of string fragmentation, while the
recombination model \cite{rec} had the other extreme view in that
the partons that are originally in the incident hadrons recombine to
form the detected hadrons. Whereas the Fritiof model \cite{fri} is a
refinement of the original Lund model, ECOMB represents an
enormous step of upgrading of the original recombination model. Most
importantly, it treats hadronization in the central region with due
regard to hadron size as well as color dynamics.

The paper is organized as follows. In Sec. 2 we review the Geometrical
Branching Model (GBM) with emphasis on the eikonal formalism for
multiparticle production. Then the parton model is incorporated into
the GBM in Sec.\ 3. The parton number distribution at each impact
parameter is used   only as the initial condition for the dynamical
evolution process of color mutation, which is discussed in Sec.\ 4. The
comparison of the results from Monte Carlo calculation with the data
on intermittency is carried out in Sec.\ 5. Some concluding remarks are
made in the final section.

\section{A Review of the Geometrical Branching Model}

The aim of GBM is to describe multiparticle production in hadronic
collisions through soft interaction \cite{gbm}. To focus on processes in
which hard subprocesses are unimportant, we confine our attention to
the energy range $10<\sqrt{s}<100$ GeV, which covers the CERN ISR
energies. The model is constructed to possess the properties of
geometrical scaling \cite{ddd} and approximate Koba-Neilsen-Olesen (KNO)
scaling
\cite{kn} that are observed in that energy range. It consists of two
parts: the geometrical part that is based on the eikonal formalism of
hadronic collisions, and the branching part that describes how
hadrons are produced. The details of the later will be modified and
improved in the following sections. Here we review the general
framework of how the two parts are put together.

In terms of the eikonal function $\Omega(b)$, which is assumed
known from an independent source, the elastic, inelastic, and total
cross sections are
\begin{eqnarray}
\sigma_{el}=\int {d^2b\, (1-e^{-\Omega(b)})^2} \quad ,
\label{2.1}
\end{eqnarray}
\begin{eqnarray}
\sigma_{inel}=\int {d^2b\, (1-e^{-2\Omega(b)})} \quad ,
\label{2.2}
\end{eqnarray}
\begin{eqnarray}
\sigma_{tot}=\int {d^2b\, 2(1-e^{-\Omega(b)})} \quad .
\label{2.3}
\end{eqnarray}
In the energy range stated above there is geometrical scaling; i.e.,
$\sigma_{el}$/$\sigma_{tot}$ is roughly constant \cite{ddd}. That
property can be guaranteed, if $\Omega$ depends only on the scaled
impact parameter $R$, where
\begin{eqnarray}
R = b/b_0(s) \quad,
\label{2.4}
\end{eqnarray}
so that (\ref{2.1})-(\ref{2.3}) may be written as
\begin{eqnarray}
\sigma_{el}=\pi b_0^2(s)\int^\infty_0 {d R^2\, (1-e^{-\Omega(R)})^2} \quad ,
\label{2.5}
\end{eqnarray}
\begin{eqnarray}
\sigma_{inel}=\pi b^2_0(s)\int^\infty_0 {d R^2\, (1-e^{-2\Omega(R)})} \quad ,
\label{2.6}
\end{eqnarray}
\begin{eqnarray}
\sigma_{tot}=\pi b^2_0(s) \int^\infty_0 {d R^2\,
2(1-e^{-\Omega(R)})} \quad .
\label{2.7}
\end{eqnarray}
In order that the inelasticity function
\begin{eqnarray} g(R)=1-e^{-2\Omega(R)}
\label{2.8}
\end{eqnarray} satisfies the normalization condition
\begin{eqnarray}
\int^\infty_0 {d R^2 g(R)}=1 \quad ,
\label{2.9}
\end{eqnarray} we are free to set the scale $b_0(s)$ by requiring
\begin{eqnarray}
\sigma_{inel}=\pi b^2_0(s)\quad .
\label{2.10}
\end{eqnarray} The function $g(R)$ describes the probability of
having an inelastic collision at $R$.

The multiplicity distribution $P_n$ is a result of sampling the
multiparticle production processes by many collisional events, each of
which may have a different impact parameter. Hence, $P_n$ should
have the form
\begin{eqnarray} P_n=\int{d R^2 g(R) {\bf Q}_n (R)} \quad ,
\label{2.11}
\end{eqnarray}
where ${\bf Q}_n(R)$ is the probability of producing $n$
particles at impact parameter $R$. Both
$P_n$ and ${\bf Q}_n$ are normalized by
\begin{eqnarray}
\sum_n P_n=\sum_n {\bf Q}_n=1 \quad .
\label{2.12}
\end{eqnarray}

The eikonal formalism is, of course, quite general. It satisfies
unitarity, which relates elastic and inelastic scattering amplitudes.
Furthermore, it emphasizes the spatial properties of the colliding
hadrons, which are known to be extended objects. Any treatment of
the collision process that ignores the geometrical aspects of the
hadrons leaves out some part of the physics of the problem, which cannot be
unimportant for soft processes.

The inelasticity function $g(R)$, defined in (\ref{2.8}), can be
expanded in a power series
\begin{eqnarray}
g(R)=\sum^\infty_{\mu=1} \pi_{\mu}(R)
\label{2.13}
\end{eqnarray}
where
\begin{eqnarray}
\pi_{\mu}(R)= {\left[2\Omega (R)\right]^{\mu}\over{\mu !}}
e^{-2\Omega (R)}.
\label{2.14}
\end{eqnarray}
The $\mu$th-order term may be regarded as the
$\mu$th-order rescattering contribution, and can be related to the
$\mu$-cut-Pomeron $\cite{ani,hp}$. In this paper we use
$\mu$-cut-Pomeron only as a terminology in reference to the
$\mu$th term in the expansion (\ref{2.13}). For $\mu\ge 2$,
$\pi_{\mu}$ is small compared to $\pi_1$ except at small $R$ for any
reasonable $\Omega (R)$. For $pp$ collisions the well-determined
form for the eikonal function is $\cite{cy}$
\begin{eqnarray}
\Omega(R)=-\ell n\left(1-0.71e^{-1.17R^2}\right) \quad,
\label{2.15}
\end{eqnarray}
which has been used to give a good description of
$d\sigma/dt$ $\cite{cy2}$. To get an estimate of the average $\mu$,
let us define
\begin{eqnarray}
\bar{\mu}(R) = \sum ^\infty_{\mu=1}
\mu\pi_{\mu}(R)/\sum^\infty_{\mu=1} \pi_{\mu}(R)
\label{2.16}
\end{eqnarray}
and
\begin{eqnarray}
\left<\mu\right>=\int{dR^2 \, g(R)\bar{\mu}(R)} =
\int{dR^2 \, 2\Omega(R)}.
\label{2.17}
\end{eqnarray}
Using (\ref{2.15}) in (\ref{2.17}), one gets
$\left<\mu\right>=1.6$ \cite{hw}. Thus unless
$R\simeq 0$, high-order terms involving $\mu > 2$ are negligible. But
for the high-$n$ behavior of the multiplicity distribution $P_n$, small
$R$ collisions are very important.

As far as the geometrical aspect of the collision problem is concerned,
the above discussion summarizes all that is important. To go further
in multiparticle production, it is necessary to model the dynamics of
soft interaction. Before we enter into the details of that, we consider
here the connection between particle production and geometry. At
each $R$, $\pi_{\mu}(R)$ gives the probability of having
$\mu$-cut-Pomerons, for which we use ${\bf B}^{\mu}_n$ to denote
the probability of producing $n$ particles. Thus the overall $P_n$ is a
convolution
\begin{eqnarray} P_n=\int
dR^2\sum^\infty_{\mu=1}\pi_{\mu}(R){\bf B}^{\mu}_n
\label{2.18}
\end{eqnarray} with $\sum_n {\bf B}^{\mu}_n=1$. Going back to the
general formula (\ref{2.11}), it is clear from (18) that ${\bf Q}_n$ is a
weighted average of ${\bf B}^{\mu}_n$:
\begin{eqnarray}
{\bf Q}_n(R)=\sum^\infty_{\mu=1}\pi_{\mu}(R){\bf
B}^\mu_n/\sum^\infty_{\mu=1}\pi_{\mu}(R),
\label{2.19}
\end{eqnarray} where (\ref{2.13}) has been used.

Using (\ref{2.14}) and (\ref{2.15}), the dependence of $\pi_{\mu}(R)$
on $R$ is shown in Fig.\ 1 for various values of $\mu$.  Evidently, at
large values of $R$ only the $\mu =1$ term is important. At $R\equiv
0$, terms up to $\mu = 6$ can still make   small contributions. The
sum in (\ref{2.13}), when truncated at $\mu = 6$, is shown by the heavy
dashed line in Fig.\ 1, and well approximates $g(R)$ shown by the
heavy solid line in Fig.\ 1. Thus we see in (\ref{2.19}) that at smaller
$R$ more terms in the sum over $\mu$ are needed, resulting in more
particles produced, as is physically reasonable.

\section{Incorporation of the Parton Model}

In an earlier version of the GBM \cite{hp}, a self-similar branching
process is assumed for ${\bf B}^\mu_n$, whose $n$ dependence is
determined by simulation. The defect of that approach is that the
dynamics of branching is {\it ad hoc}, not based on QCD. The
guiding principle is self-similarity, i.e., a cluster of mass
$m$ branches into two subclusters of masses $m_1$ and $m_2$ with
probability
$D(m, m_1, m_2)$ that is independent of any specific mass scale. The
aim is to reproduce the intermittency phenomenon observed, as the
resolution scale is varied \cite{ddk}. We now want to improve the
model by incorporating nonperturbative QCD to the extent feasible.
Branching becomes a derived property in the present approach,
instead of an assumed property in ECCO \cite{hp}.

To do so, it is necessary to put the model in a framework in
which the interactions of quarks and gluons can be examined
in detail, with hadrons being the end product of the
hadronization process. Since our aim is to study soft
interaction in low-$p_T$ production processes, rigorous
calculation from first principles is not possible. Even the use of
the parton model may be questionable when the virtuality
$Q^2$ that is relevant to the process is not high. However,
bearing in mind that Feynman's  first paper on the parton
model addresses specifically the soft interaction problem
\cite{fey}, we shall similarly frame our investigation in the
language of partons without using perturbative
approximations. Indeed, it is because the interactions are soft
that we have to follow the time evolution of the partons.

An important departure from the usual application of the parton
model is that we combine the property of hadrons being spatially
extended objects with the property of hadrons being made up of
smaller constituents. Clearly, for most collisions whose impact
parameters are nonzero, only portions of the partons in the incident
hadrons overlap and interact, so the number of hadrons produced
depends on that overlap. The event-to-event fluctuation of the
particle multiplicity therefore depends strongly on the
impact-parameter fluctuation, and the event averaged parton
distributions, as determined from the structure functions, are of no
use in that respect.

To implement our amalgamation of the two properties stated above,
we adopt the eikonal description of Sec. 2 and modify (\ref{2.19}) to
refer to parton number instead of hadron number. Thus let us define
${Q}_ \nu (R)$ to be the probability of involving $\nu$ partons in
soft interaction in the overlap region when the two incident hadrons
are separated by a scaled impact parameter $R$; let $B^{\mu}_{\nu}$ be
the probability of having $\nu$ partons in the
${\mu}$-cut Pomeron.  Note that we have
used $Q_\nu$ and $B^{\mu}_{\nu}$ in place of ${\bf Q}_n$ and
${\bf B}^\mu_n$,
when we refer to the $\nu$ partons instead of the $n$ particles. The
parton and particle distributions are, of  course, very different. As in
(\ref{2.19}), $Q_\nu$ and $B^{\mu}_{\nu}$ are related by
\begin{eqnarray} {Q}_ \nu (R) = g^{-1}(R) \sum^{\infty}_{\mu = 1}
\pi_{\mu}(R) B^{\mu}_{\nu} \quad .
\label{3.1}
\end{eqnarray} We emphasize that in (\ref{3.1}) the subscripts $\nu$
refer to the number of partons that initiate the evolution process
leading toward the hadrons produced in the central rapidity region.
The nonoverlapping portions of the incident hadrons produce the
leading particles in the fragmentation regions.
If we use ${\cal E}_{\nu\rightarrow n} \{\cdots\}$ to denote  the evolution
process that takes the $\nu$ partons to the $n$ hadrons, we may express
the relationship between $Q_\nu(R)$ and ${\bf Q}_n(R)$ symbolically as
\begin{eqnarray}
{\bf Q}_n(R) =  {\cal E}_{\nu\rightarrow n}
\{\,{Q}_\nu(R)\,\}
\quad.
\label{3.5a}
\end{eqnarray}
The description of ${\cal E}_{\nu\rightarrow n}$ is the main task of this
paper.

A way of thinking about $B^{\mu}_{\nu}$ is to consider the
$\mu = 1$ case, for which we have a one-Pomeron exchange diagram
for the elastic scattering amplitude with the Pomeron being cut to
expose the internal lines on mass shell.  If the Pomeron is represented
by a ladder of gluons with quark loops, then cutting it reveals the
gluons, quarks and antiquarks, which constitute the partons of
$B^1_{\nu}$.  Being the probability (i.e., absolute square of the amplitude
for $p+p\rightarrow \nu$ partons) of having $\nu$ partons in the one
cut-Pomeron,
$B^1_{\nu}$ is, of course, more general than any specific model
approximating that Pomeron by ladders or otherwise.  Similarly,
$B^{\mu}_{\nu}$ is the corresponding probability, when $\mu$ Pomerons
in the elastic amplitude are cut.

Since there exists no rigorous derivation of $B^{\mu}_{\nu}$,
we shall assume that it is Poisson distributed around some mean
number $\bar{\nu}$ of partons.  It is reasonable, since it is known
that a cut ladder corresponds to a multiperipheral diagram for a
$\nu$-particle production amplitude, whose rapidity distribution is
uniform, and multiplicity distribution Poissonian.   The mean number
$\bar{\nu}(\mu, s)$ can depend on both $\mu$ and $s$, since we
expect $\bar{\nu}$ to increase with $\mu$ and $s$.  Lacking any
information on that dependence, except that the $s$ dependence
should be logarithmic, this being $\log{s}$ physics, we adopt the following
parameterization
\begin{eqnarray}
\bar{\nu} (\mu, s) = \hat{\nu} (s) \mu^{a(s)} \quad ,
\label{3.2}
\end{eqnarray}
where
\begin{eqnarray}
\hat{\nu} (s) = \nu_0 + \nu_1\, {\rm ln} s \quad ,
\label{3.3}
\end{eqnarray}
\begin{eqnarray}
a(s) = a_0 + a_1\,  {\rm ln} s +a_2\, {\rm ln}^2 s \quad .
\label{3.4}
\end{eqnarray}
We shall use these in
\begin{eqnarray}
B^{\mu}_{\nu} ={1 \over \nu!} \left[\bar{\nu}
(\mu , s) \right]^{\nu} e^{-\bar{\nu} (\mu , s)} \quad ,
\label{3.5}
\end{eqnarray}
which in turn is used in (\ref{3.1}) to determine
${Q}_ \nu (R)$.

In the next section we describe the algorithm for parton evolution in
color space event by event.  For each event we simulate a scaled
impact parameter $R$.  For that $R$, we use ${Q}_ \nu (R)$ to
simulate an initial parton configuration involving $\nu$ partons.
After the completion of the evolution process, symbolized by ${\cal
E}_{\nu\rightarrow n}$,
$n$ particles are produced for that event.  Repeated simulation of many
such events results in a multiplicity distribution $P_n$ at each $s$, a
process expressed analytically by (\ref{2.11}).
 From $P_n(s)$ we can
calculate the average multiplicity
$\left<n\right>(s)$, and the standard moments
\begin{eqnarray}
C_q(s) = \left<n^q\right> /\left<n\right>^q \quad .
\label{3.6}
\end{eqnarray}
We vary the parameters in (\ref{3.3}) and (\ref{3.4})
to fit the data on $\left<n\right>(s)$ and
$C_q(s)$ for $s$ in the range $10 < \sqrt{s} < 70$ GeV.

Although the hadronization procedure has not yet been described, we
give the result here first to close our discussion on the initial parton
distribution.  We do this so as not to lose sight of the global features of
multiparticle production, when our attention is turned to the local
properties in the following sections.  The data that we want to fit are
$\left<n\right>_{\rm ch}$ and $C_q$ from Refs.\ \cite{dat}.
They are shown in Figs. 2 and 3.  The solid lines are our
calculated results using
\begin{eqnarray}
\nu_0 &=& -5.10\, ,\quad \quad \nu_1 = 4.03\, ,\nonumber\\
a_0 &=& -1.1\, ,\quad a_1 = 0.41\, , \quad a_2 = -0.025 \, ,
\label{3.6a}
\end{eqnarray} in (\ref{3.3}) and (\ref{3.4}) with $s$ in units of
GeV$^2$.  The relative $s$-independence of
$C_q(s)$, which is a manifestation of the KNO scaling, is nontrivial
when $\left<n\right>(s)$ more than doubles over the energy range
considered for soft production only.  The fits in Figs.\ 2 and 3
are evidently very satisfactory.

Before we go into the details of the color dynamics that is responsible
for the calculated results, let us show the parton number distributions
${Q}_ \nu (R)$.  They are plotted in Fig.\ 4 for four
representative values of $R$ and for $\sqrt{s}= 52$
GeV.  Note that the peak does not move below $\nu=25$, as $R$ is
increased; it is due to the fact that the minimum $\nu$ is 1, no matter
how large $R$ is.
 Note  also that the typical parton numbers are not
large by the standards of hard processes at much higher
energies.  That is connected with the nature of those partons
that initiate the evolution process, a subject we turn to next.

\section{Color Mutation}

We now consider the color dynamics of soft interaction. The initial
state is that there are $\nu$ partons distributed in some fashion in a
linear array in rapidity space. This is a consequence of the parton
model in that the partons are in the incident hadrons to begin with,
and the collision rearranges those partons and sets off the evolution
process that takes those partons to the final state, where the produced
hadrons are decoupled. The evolution process involves quarks and
antiquarks emitting and absorbing gluons. Since the process is not
perturbative, there is no analytical method to track the time development
of the process. Thus we shall use Monte Carlo simulation to generate
the configuration at each time step.

There are two spaces in which we must track the motion of the color
charges. One is the two-dimensional color space; the other is the
one-dimensional rapidity space. The latter can be extended to include
the azimuthal angle
$\phi$ and the transverse momentum $\left|\vec{p}_T\right|$, but in
our first attempt here we integrate over those variables and examine
the simpler problem of a 1D system. As for the color space, it is 2D,
and can be spanned by the ($I_3$,$Y$) axes, as in $SU(3)$ flavor. The
configuration in the two spaces are coordinated in the following sense.
Starting from the extreme left end of the rapidity ($\eta$) space, for
which the system that contains no partons is by definition color neutral
and therefore is
represented by a point at the origin of the  color space, we move in
the positive direction in $\eta$ space until we cross a color
charge. At the point to the right of that color charge, the
corresponding point in the color space jumps from the origin
to a position that represents the color of the charge that has
been crossed. Let us denote that jump by a vector. Then each
time we cross a color charge in the
 $\eta$ space, there is a corresponding vector added to the previous
point in the color space. The succession of additions of those vectors,
each one starting from the tip of the previous one, forms a path. Since
the whole parton system is color neutral, by the time we have
moved to the extreme right in the $\eta$ space, the path in the color
space returns to the origin, thus forming a closed loop. Such a loop
may be self-intersecting at various points. We call such a closed path
a configuration of the system in color space.

The essence of our evolution process is to track the configuration of
the system at each time step. The configuration  changes because the
colors of the partons can mutate, as they interact through the
emission and absorption of gluons. The way that each color
charge mutates is determined by the use of an energy
principle, which is proposed on the grounds that the system
attempts to lower its energy by changing its color
configuration, but fluctuating forces at the microscopic level
cause the changes to be erratic. At the same time, the
confinement force tends to reduce the distances among the
color charges, which are separated initially because of the
initial momenta that the partons possess, while being
constituents of the incident hadrons. Since we work in the 1D
$\eta$ space and cannot use perturbative QCD, we do not treat the
scattering problem among the partons in either the momentum or the
coordinate space. The evolution of the configuration in the color space
is what is important.

When the closed path in the color space evolves to a configuration in
which the path crosses the origin, a point in the evolution is then
reached that requires special attention. At that point, the
configuration may be regarded as having two closed paths,
corresponding to two color neutral subsystems. We assume that a
subsystem with no net color charge (i.e. $I_3$ and $Y=0$ in color
space) is in the lowest $SU(3)$ representation, that is, a singlet.
This assumption is based mainly on the expectation that a
nonsinglet system requires more energy to sustain it than
a singlet system.  Relaxation to the lowest energy state,
despite fluctuations, is the principle that guides the
evolution.  Since two spatially separate singlet subsystems do
not interact, we regard a branching process to have taken
place: the original large cluster of partons has partitioned
itself into two smaller color neutral subclusters.  Thereafter,
we repeat our procedure of color mutation, but now applied to
the two subclusters separately. This process is repeated again
and again until all subsystems consist of only quark-antiquark
pairs that cannot branch any further. However, a $q\bar{q}$
subsystem can be a resonance, in which case it can decay into
pions and kaons.

So far, we have described only qualitatively the evolution process of
color mutation and branching. The quantitative implementation of the
procedure will be given below. The conceptual basis of our approach
to soft production incorporates many features of strong interaction
that are generally regarded as being reasonable and physical, but
lacking calculational schemes with results for comparison with data.
We now amend that defect. The procedure that we adopt to carry out
the calculations is not unique, and can be modified and improved in
time. But the framework of  ECOMB is fixed by the combination of
eikonalism, color mutation, and branching. Our present procedure is as
follows.

\subsection*{A. Initial Distributions of the Partons}

As described in Sec.3, we start out with $\nu$ partons for a given
event. It is a number generated by use of the eikonal formalism with
variations in the impact parameter taken into account. How the $\nu$
partons are distributed in rapidity is drawn from the parton model. An
important basis of our model is that partons are in the incident hadrons
before the collision, and are not to be regarded as $q\bar q$ pairs that are
excited from the vacuum due to the stretching of strings in the string
model. Thus our partons have a momentum-fraction distribution as
determined in leptoproduction, but with $Q^2$ as small as can be achieved.

In the parton model there has always been a question of how gluons
hadronize in low-$p_T$ processes, when the produced particles are
predominantly pions. In the recombination model for particle
production in the fragmentization region \cite{rec}, the problem is
handled by saturating the sea, i.e., by converting all gluons to quarks
and antiquarks at the outset. The normalization of the hadronic
inclusive cross section turns out to be correct, when the recombination
mechanism is applied to all quarks and antiquarks that are nearby in
phase space. We now use the same saturation of the sea in our
problem here, as we attempt to treat the central region. Thus when we
refer to partons in the following, we shall mean only quarks and
antiquarks. The $\nu$ partons are therefore color triplets and
antitriplets of equal numbers so that collectively they form a color
neutral system. Their distributions in rapidity and color spaces are to
be considered separately.

\subsubsection*{(1) Distribution in rapidity space}

By rapidity we mean space-time rapidity $\eta$, where
\begin{eqnarray}
\eta = {1\over 2}\, \ell n\, {t + z \over t - z} \quad .
\label{4.1}
\end{eqnarray}
Momentum rapidity $y$ will not be used throughout
the calculation until the end, when the momentum  of a produced
particle is to be determined. We use the $\eta$ variable because the
spatial separations between partons change during the evolution process
due to the color forces.

The usual parton distribution in momentum fraction (at the lowest accessible
$Q^2$) implies roughly a flat $\eta$ distribution with rapid
damping toward zero at large $\eta$. We approximate that distribution for
the forward-going incident hadron by
\begin{eqnarray}
\rho^{(f)}(\eta) =
\rho_0\, , \hspace{1.2in} 0 \le \eta \le
\eta_c\,  ,\hspace{1.5cm}\nonumber\\
\rho_0 \, {\eta_{\rm max} - \eta \over
\eta_{\rm max}- \eta_c}\,
,\quad \hspace{.25in}\eta_c \le \eta
\le \eta_{\rm max}\, ,\hspace{.5in}
\label{4.2a}
\end{eqnarray}
where $\rho_0=\nu/(\eta_{\rm max}+\eta_c)$,
which corresponds to a total of $\nu/2$ partons when integrated over
$\eta>0$. For the partons belonging to the backward-going incident hadron,
the
$\eta$-distribution is
\begin{eqnarray}
\rho^{(b)}(\eta) =
\rho_0\, , \hspace{1.2in} -\eta_c \le \eta \le 0
\,  ,\hspace{1.5cm}\nonumber\\
\rho_0 \, {\eta_{\rm max} + \eta \over
\eta_{\rm max}- \eta_c}\,
,\quad \hspace{.25in}-\eta_{\rm max} \le \eta
\le -\eta_c\, .\hspace{.5in}
\label{4.2b}
\end{eqnarray}
The partons in each of  $\rho^{(f)}$ and $\rho^{(b)}$ should add up to be
color neutral. The overall parton distribution in the initial state is then
\begin{eqnarray}
\rho(\eta)=\rho^{(f)}(\eta)+\rho^{(b)}(\eta)\, .
\label{4.2}
\end{eqnarray}
 For each event, the
partons are, to first approximation, distributed randomly in $\eta$
according to $\rho(\eta)$.

The above distribution describes the rapidities of the partons before the
collision of the incident hadrons. Upon collision those partons  interact,
even if only softly. The conventional wisdom about soft interaction is that
the range of interaction in rapidity is short. That follows from the known
fact that the correlation between produced particles is short-ranged. Thus
apart from a limited range
$(\, -\eta_0 \le \eta \le \eta_0\, )$ in the central region where the partons
from the two incident hadrons interact, most partons outside that range
are undisturbed by the collision.  We describe the effect of the interaction
by modifying the initial distributions $\rho^{(f)}(\eta)$ and
$\rho^{(b)}(\eta)$ in that interaction region to $\rho^{(1)}(\eta)$ and
$\rho^{(2)}(\eta)$, respectively, where
\begin{eqnarray}
\rho^{(1)}(\eta) = \rho_0{\eta + \eta_0 \over 2 \eta_0 }\,
\theta(\eta+\eta_0)\, \theta(\eta_0 - \eta) + \rho^{(f)}(\eta)\,
\theta(\eta-\eta_0) ,
\label{29.2}
\end{eqnarray}
\begin{eqnarray}
\rho^{(2)}(\eta) = \rho_0{-\eta + \eta_0 \over 2 \eta_0 }\,
\theta(\eta+\eta_0)\, \theta(\eta_0 - \eta) + \rho^{(b)}(\eta)\,
\theta(-\eta-\eta_0) .
\label{29.3}
\end{eqnarray}
Note that the total parton distribution is unchanged, i.e.,
\begin{eqnarray}
\rho^{(1)}(\eta) + \rho^{(2)}(\eta) = \rho(\eta)\, .
\label{29.4}
\end{eqnarray}
The shapes of $\rho^{(1)}(\eta)$ and $\rho^{(2)}(\eta)$ are now
trapezoidal: some of the partons initially between $\eta=0$ and
$\eta_0$  in
$\rho^{(f)}(\eta)$ are dragged to the negative $\eta$ region, and some
between
$\eta_0$ and 0 in
$\rho^{(b)}(\eta)$ are dragged to the positive $\eta$ region.   Fig.\
5 illustrates how the distributions look like at a low energy where
the trapezoids are almost triangular. This rearrangement may not seem
 significant for the inclusive cross section of the produced hadrons,
since $\rho(\eta)$ is unchanged, but it turns out to be important for
multiplicity fluctuations in smaller bins. The point is that $\rho^{(f)}$ and
$\rho^{(b)}$ are two color-neutral clusters, coming from the two incident
hadrons.  The rearrangement caused by the collision creates an
overlapping region so that there is no smaller segment within the entire
range,
$-\eta_{\rm max} \le \eta \le \eta_{\rm max}$, where one can find color
neutrality, except by accident. The  color mutation process is sensitive to the
fact that the partons distributed according to $\rho^{(1)}(\eta) +
\rho^{(2)}(\eta)$ has no color-neutral subclusters, so that the evolution
must begin from the whole cluster, not from the two neutral subclusters,
$\rho^{(f)}$ and $\rho^{(b)}$, separately.

One can find a rough parallel between our $\rho^{(1,2)}(\eta)$ and the
strings in the Fritiof model \cite{fri} or in the dual parton model
\cite{dpm}.  However,  a string in DPM is stretched between a quark in one
incident proton with a diquark in another proton. Our
$\rho^{(1)}(\eta)$ is a color-neutral cluster of partons distributed from
$-\eta_0$ to $\eta_{\rm max}$.

To summarize, whereas $\rho^{(f)}(\eta)$ and $\rho^{(b)}(\eta)$ are the
initial parton distributions of the incident hadrons, $\rho^{(1)}(\eta)$ and
$\rho^{(2)}(\eta)$ are the parton distributions after the hadrons pass
through the interaction region.  The  evolution of the color dynamics
begins thereafter starting with $\rho^{(1)}(\eta)$ and
$\rho^{(2)}(\eta)$.

\subsubsection*{(2) Distribution in color space}

In the color space spanned by $I_3$ and $Y$, a quark is represented
by a vector, which has the coordinates of one of the triplets: $(1/2,
1/3), (-1/2, 1/3)$, and $(0,\,-2/3)$. An antiquark is
represented by a vector directed opposite to one of the above.
As a distribution of partons in the $\eta$ space is generated, say, $
\rho^{(2)}(\eta)$ between $-\eta_{\rm max} \le \eta \le \eta_0$,  we assign
to each parton a color vector
consistent with the requirement that the quarks and antiquarks come in
pairs, but their
orderings in the $\eta$ and the color spaces are totally random.
The partons in $\rho^{(1)}(\eta)$ are distributed similarly, but completely
independently, between $-\eta_0 \le \eta \le \eta_{\rm max}$. The total
distribution $\rho(\eta)$ in $-\eta_{\rm max} \le \eta \le \eta_{\rm
max}$ is the sum of these two sets, whose partons are merged in the
$\eta$ and the color spaces.
 Thus as we move from the extreme left in  the
$\eta$ space in the $+\eta$ direction, we begin at the origin of the color
space, and, each time when we move over a parton in the $\eta$
space, we add a vector (corresponding to its color) with its tail
to the head of the previous vector (or to the origin if it is the
first parton). A succession of such vectors forms a trajectory in
the color space. The trajectory eventually ends at the origin,
thus forming a closed path, when $\eta$ reaches the extreme
right.

Note that the path in the color space is dual to the ``path'' in the 1D
$\eta$ space in the sense that the segment between two partons in
the $\eta$ space is mapped to a point in the color space, while the
point where a parton is located is mapped to a segment of the path in
the color space. Thus the closed path is made up of $\nu$ short
segments. An example of a path in color space is shown in Fig.\ 6. The
sequence of color charges represented by the vectors is indicated by
$r\bar{b}rg \cdots \bar{b}r$ in the figure. We have used the
notation
$r=(1/2, 1/3)$, $g=(-1/2, 1/3)$, and $b=(0,\, -2/3)$. The origin
is labeled by 0.

\subsection*{B. Evolution in the color and rapidity spaces}

We now consider the evolution of a configuration due to QCD
dynamics. A nonperturbative treatment of $\nu$ simultaneously
interacting color charges is, of course, too difficult to contemplate here.
We reduce the problem by considering pairwise near-neighbor
interaction via the exchange of a gluon in any of the $s$-, $t$-, or
$u$-channel, whichever is applicable. Starting from the extreme
left in the $\eta$ space, we regard the ordered chain of $\nu$
partons as having $\nu-1$ links (with varying link lengths). Its
dual path in the color space has $\nu$ vectors, meeting at
$\nu-1$ vertices, not counting the endpoints, which are located
at the origin. Pairwise near-neighbor interaction means that we
consider the $\nu-1$ links  in $\eta$ space one at a time, according to a
rule to be specified below. After the interactions at all $\nu-1$ links
are considered, the evolution of the whole configuration is
regarded as having taken one time step, and the process is then
repeated.

\subsubsection*{(1) Color interaction}

Consider any link in the chain, i.e., a pair of partons. Suppose that the
pair is ($r,\bar{g}$). The only possible outcome of an interaction by
the emission and absorption of a gluon, besides remaining as
($r,\bar{g}$), is ($\bar{g}, r$), since the sum of the two color vectors
must remain invariant. If the pair consists of quarks, such as ($r,g$),
the only outcome of an interaction is again either ($r,g$) or ($g,r$).
Only in the case of a color-neutral pair, such as ($r,\bar{r}$), can the
outcome be any such pair, i.e., ($r,\bar{r}$), ($g,\bar{g}$),
($b,\bar{b}$) or their exchanged pairings. To determine the specific
outcome of an interaction, we use a statistical method.

For every global configuration $\alpha$ of the $\nu$ partons, there is
an associated energy $E_{\alpha}$. How
$E_{\alpha}$ is determined will be described in the following
subsection. Whenever a local pair of partons interact, the outcome
may or may not affect the global configuration. Let $c$ be the total
number of possible configurations. From the foregoing discussion $c$
can only be 2 or 6, the latter being for a color-neutral pair. Our
statistical approach to the determination of the global configuration
$\alpha$, consistent with favoring the lowest energy state, requires that
the probability for configuration $\alpha$ to occur is
\begin{eqnarray} P_{\alpha}=e^{-\beta E_{\alpha}}/Z, \qquad
Z =
\sum^c_{\alpha=1}e^{-\beta E_{\alpha}},
\label{4.3}
\end{eqnarray} where $\beta$ is a free parameter.

Using the Metropolis algorithm, we use $P_{\alpha}$ to determine the
outcome of a local interaction at every link. The result affects only one
vertex of the closed path in the color space, one vertex at a time. As
we go down the chain, every vertex is then tested for possible
changes. At the end of one complete sweep for one time step, the
whole closed path has undergone a color mutation, resulting in a new
path. Note that our procedure  incorporates several features of the
dynamics governing the system. Locally, there is the QCD dynamics of
gluon exchange between quarks and antiquarks. The outcome of the
local interaction depends on the global configuration in the color
space. The statistical treatment takes into
account the fluctuating nature of the many-body problem.

\subsubsection*{(2) Energy of a color configuration}

The color mutation process is an attempt by the dynamical system to
relax to its lowest energy state. It starts initially at a high energy state
because the collision puts the partons in a spatially spread-out
domain by virtue of the initial momenta of the partons that are essentially
uninterrupted by the soft collision, i.e., no large momentum transfers.
Color charges that are dispersed in space have high potential energy.
Since soft interactions do not result in significant changes in the
parton momenta, the energy of the system is controlled mainly by the
potential energies residing at all the links in the $\nu$-parton chain.

Specifically, consider the $i$th link in a chain, where
$i= 1, \,2, \cdots ,  \nu -1$.  The net color charge that
the link sees on the left side is
\begin{eqnarray}
\vec{C}_i=\sum^i_{j=1}\vec{c}_j \quad ,
\label{4.5}
\end{eqnarray}
where $\vec{c}_j$ is the vector in color space
representing the color of the $j$th parton. Due to the color neutrality
of the whole system, the net color that the link sees on the right side
is $-\vec{C}_i$. Since the potential energy
at the link $i$ is proportional to the net  color charges on the two sides
of the
link,  we write the energy
$E_i$ of the
$i$th link in the simple form
\begin{eqnarray}
E_i=\left|\vec{C}_i\right|^2 \quad ,
\label{4.6}
\end{eqnarray}
where the proportionality factor is
absent for the following reason. The confinement potential should be
proportional to the distance between color charges. Here $\vec C_i$ is the
total color of all the partons to the left of the $i$th link, so the relevant
effective distance should be measured from the center of those partons on
the left side to the center of the partons on the right side. That is a
global distance that is roughly half the total $\eta$ range, insensitive
to the position of the $i$th link. Since $E_\alpha$ is never used
without $\beta$ in (\ref{4.3}), the omission of the proportionality
factor in (\ref{4.6}) is equivalent to letting it be absorbed in the
definition of $\beta$.
 The energy associated with the whole configuration
$\alpha$ is
\begin{eqnarray}
 E_{\alpha}=\sum^{\nu -1}_{i=1} E_i \quad ,
\label{4.7}
\end{eqnarray}
where the sum obviously depends on the particular
path in color space that the configuration $\alpha$ takes.

It is this energy $E_{\alpha}$ in (\ref{4.7}) that the system attempts to
lower by color mutation, as it changes from one configuration
$\alpha$ to another, $\alpha^{\prime}$. Because of the additive nature of
$E_{\alpha}$ in (\ref{4.7}), the probability of mutation is determined
by the local energy difference, as can be seen in the following example.
Consider a link for which the number of possible configurations is two, i.e.,
$c=2$ in (\ref{4.3}). Then
$P_{\alpha}$ can be written as
\begin{eqnarray}
P_\alpha = {1\over 1+e^{-\beta\, \Delta E}}\, ,
\label{4.7a}
\end{eqnarray}
where $\Delta E$ is the change of energy in going to the new configuration
$c'$
\begin{eqnarray}
 \Delta E=E_{\alpha^{\prime}}-E_\alpha=E^{\prime}_i-E_i\, .
\label{4.7b}
\end{eqnarray}
$E^{\prime}_i$ is the energy at the $i$th link in the new configuration.
Note that $\Delta E$ is now locally determined at the link $i$. However, the
energy $E_i$ and $E'_i$ depend on the global configurations in the color
space, due to (\ref{4.5}) and (\ref{4.6}).

Since the net color charge of all the partons on one side of a link can be
large,
$E_i$ can also be large, according to (\ref{4.6}). Thus unless $\beta$ is a
small number, (\ref{4.7a}) can give a probability distribution $P_{\alpha}$
that can have an abrupt change at $\Delta E=0$. To round off the edges,
$\beta$ would have to be quite small.

\subsubsection*{(3) Spatial fluctuations}

The length of the $i$th link in the $\eta$ space is
\begin{eqnarray}
d_i=\eta_{i+1}-\eta_i \quad,
\label{4.4}
\end{eqnarray}
where $\eta_i$ is the rapidity of the $i$th parton,
counting from the left at $\eta=-\eta_{\rm max}$.
For a spatially extended color-neutral system, we have positive link
lengths, $d_i\ge 0$. We put $d_i=0$, when two partons are in the
same $\eta$ bin. The rapidity range,
$-\eta_{\rm max}\le\eta\le\eta_{\rm max}$, is divided into 256 bins in our
numerical calculation, so the smallest bin size is $2\eta_{\rm max}/256$,
which, of course, depends on the collision energy. As far as the
problem regarding the spatial distribution is concerned, we shall
assume energy independence and work only in the space that has 256
basic units, whatever $\eta_{\rm max}$ is. We denote the width of the
basic unit by
$\delta\eta$.

On the basis of  (\ref{4.3}) and (\ref{4.6}) the system undergoes color
mutation, as it evolves toward the final state.  In addition to the fluctuation
of the color charge $\vec C_i$, the length of the $i$th link, $d_i$, which is
an integer in units of $\delta \eta$, can also fluctuate.  Whether the link
length contracts or expands depends on the attractiveness or repulsiveness
of the net color forces that act on the two ends of the link.  To determine the
nature of that force is beyond the scope of this treatment.  We shall model
the change in link length by a stochastic approach, consistent with how we
have handled the color mutation part of the dynamics of the complex
system.   We allow $d_i$ to change by an integer $m_i$, i.e., $d_i \rightarrow
d_i+m_i$, where the probability for $m_i$ is specified by a distribution
${\cal P}_i(m)$.
Apart from the consideration that a contraction cannot render $d_i$
negative, we would set ${\cal P}_i(m)$ to be a uniform distribution from
$-m_-$ to $m_+$, where $m_{\pm}\ge 0$. However , if $m_->d_i$, we
truncate
${\cal P}_i(m)$  at
$m=-d_i$ and place the probabilities for contraction from $-d_i$ to $-m_-$
at
$m=-d_i$. Thus our two-parameter formula for ${\cal P}_i(m)$ can be
rewritten in the form that has
 a constant term plus a Kronecker delta term  at $-d_i$. That is, with
$\gamma_1$ and $\gamma_2$ as the two revised parameters, we have
\begin{eqnarray}
{\cal P}_i(m) = \gamma_1\, \theta(m+d_i)\, \theta(\gamma_2-m)
+[1-\gamma_1(\gamma_2+d_i)]\, \delta_{m,-d_i} \quad,
\label{4.7c}
\end{eqnarray}
which satisfies $\sum_{m=-d_i}^{\gamma_2}{\cal P}_i(m)=1.$
This distribution allows $m$ to be positive (expansion) uniformly up to
$\gamma_2$, and negative (contraction) down to $-d_i$.  The net
probability for expansion is
$\gamma_1\gamma_2$, and for contraction is $1-\gamma_1\gamma_2$.

Confinement implies that there should be a net contraction between the
subcluster of charge $\vec C_i$ and the opposite subcluster of charge
$-\vec C_i$. However, these two subclusters are at an effective distance
much larger than $d_i$, which is the length of the link specifying the
separation between the boundaries of the two subclusters. At the $i$th
link, there are microscopic color forces that can be attractive or repulsive.
${\cal P}_i(m)$ is a two-parameter description of that stochastic force,
which should lead to a net macroscopic contraction for the whole cluster.
That property is related to the value $\gamma_1\gamma_2$, which
should be less than 0.5, if contraction dominates over expansion. Of course,
this change of $d_i$ should be tested at all links in the neutral cluster
before a net effect is known.

\subsection*{C. Branching}

As we have discussed in the introduction of this section, the evolution
process described here is a branching process. A fission of the
color-singlet cluster occurs, when it contains two color-singlet
subclusters, since no confining force exists between them. The
partitioning takes place when the closed path of the mother cluster
evolves to a configuration where the path passes through the origin in
the color space so that there are two color neutral subsystems.
Thereafter, the color mutation process is applied to the two
subsystems separately and independently. This is repeated again and
again until all subclusters consist of only $q\bar{q}$ pairs.

During color mutation, it is possible that a baryon-like singlet can be
formed with a $(r,g,b)$ closed loop. Since baryon-antibaryon pair
production in $pp$ collision is rare at energies where soft
interaction is dominant, we shall ignore 3$q$ and 3$\bar{q}$
singlets, (since mass inhibition is not explicitly taken into
account), and require color mutation to continue until only
mesons are produced. If in the middle of a color chain there is
a color neutral $q\bar{q}$ pair, such as $r\bar{r}$, it is not
allowed to break off from the chain; it must continue to
mutate until a genuine neutral subcluster, counting from the
extreme left or right in the $\eta$ space, is formed.

In Fig.\ 7 we show an example of how 24 partons evolve under color
mutation, and how branching takes place. Thirty-two time steps are taken
to get to the final state of $q\bar q$ subclusters only, when hadronization
occurs.

To gain some insight on how long a typical  evolution process takes for a
cluster to partition into two, we show in Fig.\ 8(a) the distribution in
time steps for $10^3$ initial clusters each consisting of ten pairs of
$q\bar{q}$ with random color ordering. The average is about 3.5 steps for
just one branching. For the evolution processes to complete, the daughter
clusters must continue to branch successively, and the overall distribution
at the end of the evolution is shown in Fig.\ 8(b). Note that on the
average quite a large number of time steps is required for the
evolution of an average-sized cluster, much larger than what is
usually involved in a hard porcess where high virtuality is degraded
by branching.

At every stage of the evolution process, a cluster shrinks due to overall
spatial contraction. We always keep the center of the cluster invariant
to conserve momentum. When a branching occurs, the two daughter
clusters will have their own respective centers, which will remain
invariant during contraction, until they themselves branch. At the end
when a hadron is formed from a $q\bar{q}$ pair, the hadron
momentum rapidity $y$ will be identified with a value of
$\eta$ taken randomly between the $\eta$ values of the quark and
antiquark. The reason for doing this is that without knowing the mass
and the transverse momentum $p_T$ of the hadron, a precise
determination of its $y$ is not possible, nor meaningful. A random
value within a small range is good enough. A rough identification of $y$
with
$\eta$ is justified for free particles at high energy. We do not expect
that our
measure of fluctuations in the final result will be sensitive to this
$y$-$\eta$ identification.

\subsection*{D. Hadronization}

The branching process terminates, when all the neutral subclusters
are reduced to the composition of $q\bar{q}$ pairs only. Those
subclusters are identified as hadrons, which may be pions, kaons, or
resonances. Those resonances must be allowed to decay before the
total number and distribution of particles are counted for the final
state of the event. The probabilities of producing various resonances
and stable (in strong interaction) particles have been studied
experimentally in $pp$ collisions in \cite{res}. We use that reference
as a generic guide for the proportions of all particles produced in any
general hadronic collision.

In our simulation we use that guide to determine whether a neutral
subcluster is a resonance or not. In either case we give it a transverse
momentum according to a specified distribution. Let us use $k_T$ to
denote the transverse momentum of a $q\bar{q}$ pair. The
distribution we use is
\begin{eqnarray}
{dn \over dk^2_T}=
{1 \over \left<k^2_T\right>}e^{-k^2_T / \left<k^2_T\right>}
\quad ,
\label{4.8}
\end{eqnarray}
where $\left<k^2_T\right>$ is a parameter to be varied. If the
$q\bar{q}$ pair forms a pion or a kaon, then that $k_T$ becomes
the
$p_T$ of the particle without any change. However, if the
$q\bar{q}$ pair forms a resonance, then we assume an isotropic
decay distribution in the rest frame of the resonance. The
azimuthal angle
$\phi$ of $\vec{k}$ is assigned randomly. After boosting back to
the cm system, the 3-momenta $\vec{p}$ of the decay particles
are then determined. The collection of all the final-state particles results in
the exclusive distribution in the $(y,\vec p_T)$ space for the event
generated.

The charges of the particles are chosen randomly subject to the
constraint that the total charge can only be either 0, 1, or 2, on the
grounds that the leading particles in a $pp$ collision can be doubly-,
singly-, or un-charged. The particles that we treat are all the produced
particles, excluding the leading particles. Once the charge of a specific
particle or resonance is generated, the decay of the resonance is then
assigned the usual branching fractions into different channels according to
the particle data book.

\section{Results}
In the previous section we have introduced seven parameters:
$\eta_c$ and  $\eta_{\rm max}$ in (\ref{4.2a}) and (\ref{4.2b}), $\eta_0$ in
(\ref{29.2}) and (\ref{29.3}), $\beta$ in (\ref{4.3}),
$\gamma_1$ and $\gamma_2$ in (\ref{4.7c}),
and
$\left< k^2_T\right>$ in (\ref{4.8}).  They are to be varied to fit the data on
inclusive distributions and on fluctuations of the exclusive distributions.
The properties of fluctuations will be quantified in terms of the normalized
factorial moments, $F_q(\delta)$ for farious bin sizes $\delta$.

The parameters  $\eta_c$, $\eta_{\rm max}$, and  $\left< k^2_T\right>$
are essentially kinematical; they set the boundaries of the phase space in
which the partons are placed.  They do not affect the dynamics of
color mutation and the spatial fluctuation of the clusters.  The
data used to determine them are the rapidity distribution $dn/dy$
\cite{thom} and the
transverse mometum distribution $g(p_T)$ of the produced
particles in the final state.  The energy range of the data in \cite{thom}
is $22 < \sqrt{s} < 63$   GeV, for which hard scattering is negligible.
The values of $\eta_c$ and $\eta_{\rm max}$ depend on $s$. To fit the
data, we have used the values given in Table I. The value of
$\left( \left< k^2_T\right> \right)^{1/2}$ is chosen to be 400 MeV.
In Fig.\ 9 we show the rapidity distributions where the
histograms are the results of our calculation, while the data are from
\cite{thom}.  There, $\eta$ refers to pseudaorapidity, not space-time
rapidity.  The agreement is clearly satisfactory. Although in carrying
out our calculation the other parameters need to be specified also, we
have not given them in Table I because the results on
$dn/d\eta$ are insensitive to them. They will be given below.

\begin{center}
\begin{tabular}{|c|c|c|c|c|c|c|c|c|c}
\hline
$\sqrt{s}$ & 22.0 & 23.6 & 30.8 & 45.2 & 53.2 & 63.2  \\
\hline
$\eta_c$   & 1.76  & 1.55   & 1.42   & 0.889   & 0.762 & 0.635  \\
\hline
$\eta_{max}$ & 5.0 & 6.6 & 6.6  & 6.5  & 6.5 & 6.5 \\
\hline
\end{tabular}
{\\TABLE 1. The parameters for the initial condition of partons in the evolution }
\end{center}

 The multiplicity distribution at $\sqrt s =22$ GeV is shown in Fig.\ 10
for $|y|<2.5$. The calculated result evidently agrees very well with the
data \cite{ada}, especially at the high $n$ end. At the low end, there is
a slight discrepency, which could very well be due to the fact that we
have not included diffractive production. That portion of the
distribution will have negligible effect on our calculation of the
factorial moments.  We have not included in Fig.\ 10 the $P_n$ for
narrow rapidity windows, since essentially the same physical content
will be conveyed by the $F_q(\delta)$ to be presented below.  We have
not considered the UA1 and UA5 data at higher energies because
those data contain significant contributions from hard subprocesses.
The inclusion of such perturbative processes is not difficult
\cite{ch,wang}, and should be investigated in order to extend the
energy range where ECOMB can be applied.

The  parameters that influence the evolution process of
mutation and branching are $\eta_0$ , $\beta$, $\gamma_1$ and
$\gamma_2$. $\eta_0$ specifies the overlap region of the initial
color-neutral clusters,
$\beta$ pertains to the probability of color mutation, and
$\gamma_1$ and $\gamma_2$ characterize the fluctuations of the link
lengths. The data used to determine them are on factorial moments and
their variants
\cite{ddk}. It is here that we can underline the importance of
those data on fluctuations, without which we have no guidance
on how to restrict the detail dynamics of particle production.
Putting that in another way, in the absence of a procedure to
calculate from first principles, any model that fails to fit the
fluctuation data is missing some aspect of the basic dynamics. To
the extent of our awareness, very few models on soft interaction
have been put to the test of confronting those data on the factorial
moments for varying bin sizes.

The normalized factorial moments, first suggested by Bia\l as
and Peschanski \cite{bp}, can be expressed as
\begin{eqnarray}
F_q={1\over M}\sum_k{{\left<n(n-1)...(n-q+1)\right>_k\over
\left<n\right>_k^q}}
\quad ,
\label{5.1}
\end{eqnarray}
where $n$ is the multiplicity in a bin of size $\delta$,  the
average is performed over all events for the $k$th bin, and the summation
is  over all bins of the same $\delta$, $M$ being the total number of bins
$(2Y/\delta)$. Data on
$F_q$ for soft interaction have been obtained by NA22 \cite{na2}.  In
recent years Bose-Einstein correlation between identical particles has been
recognized as having an important effect on the intermittency
data at very  small $\delta$ \cite{ddk}.  However,
at this stage of our development of ECOMB, we have  not
 included BE correlation, since it is unrelated to the color
dynamics that we attempt to tune. We therefore limit our scope in this
first attempt and  focus here on $F_q$, for which we fit  the  data  of
\cite{na2} as functions of bin width, rather than on correlation functions at
very small momentum-difference squared.

The parameters that we have varied to give the best fit of the
intermittency data at ${\sqrt s}=22$ GeV are:
\begin{eqnarray}
\hat \nu(s) = 9.1, \qquad a(s) = 0.63, \\
\eta_{\rm max}=5,\qquad \eta_c=3.5,\qquad \eta_0=1.9, \\
\beta=0.0015,\qquad \gamma_1=0.077, \qquad \gamma_2=5.
\label{5.2}
\end{eqnarray}
Note that $\gamma_1\gamma_2=0.385$, implying that it is roughly twice
as likely for a link to contract than to expand. To explain why the first four
parameters listed above are different from the ones used previously for the
ISR data (in Sec.\ 3 and Table I), we remark that we have not been
able to derive from the values of
$F_q$ given by the data of NA22
\cite{na2} the values of $C_q$ that can agree with those of ISR
\cite{dat}. Thus while
the parameters in Table I are adequate for the  ISR data that
range up to 63 GeV, we must use a slightly different set of parameters at
$\sqrt s = 22$ GeV for the NA22 data that provide the only information on
intermittency.

	In Fig.\ 11 we show the intermittency data of \cite{na2}, which to our
knowledge have not been reproduced by any model, except ECCO
\cite{hp,ddk}.  The lines shown
 in Fig.\ 11 are the result of our present calculation in ECOMB.
 Evidently, the agreement between our results and the data  are
satisfactory for all values of
$\delta y$ and $q$.
It should be noted that to achieve the fits attained is highly nontrivial. If
any part of the dynamical process in generating the hadrons is altered,
one would not be able to obtain the rising factorial moments, no matter
how many parameters are used.  By working with the many parts of our
model, all of which affect the determination of $F_q$, we have
gained confidence in regarding the dynamics of color mutation  and
branching as having captured the essential properties of soft interaction.

\section{Conclusion}

It is rather satisfying that we have been able to reproduce the
intermittency data in Fig.\ 11. Scaling behavior of that type
implies self-similarity in the dynamics of particle production.
The branching process of color-neutral clusters, successively
partitioning into small clusters, is self-similar, since the same
algorithm is used at every step. Thus from the outset ECOMB has a
chance of generating a scaling behavior, in contrast to the string
model or dual parton model, which have been shown by NA22 not to
fit the intermittency data \cite{na2}.  However, to have
self-similar dynamics does not guarantee that the observed scaling
behavior can be obtained. An appropriate balance of contraction
and expansion of the clusters during the evolution is necessary to
capture the essence of the color dynamics of a complex system.
Furthermore, the initial state of the parton system at the start
of the evolution process is also important, since it can dictate
the bin size from which the scaling behavior commences. These
physical aspects of the model must be incorporated properly in
order to produce the result shown in Fig.\ 11.

We have not considered the intermittency data of UA1 \cite{ua1}
because they contain minijets. If one is willing to adopt the
point of view that such semihard subprocesses are not dominant
enough to alter the nature of intermittency, one could attempt to
apply ECOMB to the UA1 data. However, it is well known that
geometrical scaling and KNO scaling are violated at the SPS
energies, so some modification is necessary.

The most important omission in this paper is Bose-Einstein
correlation, without which our model is not completely realistic.
That defect must be corrected in an improved version. At this
stage we can only offer this version as a first step in the right
direction. Whereas the details of our parametrization here may
alter after the BE correlation is incorporated, the branching
dynamics of color mutation will hold as the basic framework of the
model. Since the power-law behavior of $F_q$ has been observed for
nonidentical particles \cite{aga}, we know that self-similarity is
necessary in the absence of BE correlation. After the
symmetrization of identical-particle states is performed, the
strength of intermittency may be enhanced. The parameters may
therefore have to be readjusted, but the main features of the
dynamics need no modification.

We have amalgamated many concepts that form various elements of
the conventional wisdom about soft interaction. They include: (a)
hadrons having sizes, (b) eikonalism, (c) parton model, (d)
interaction of quarks via gluons, (e) statistical properties of a
many-body system, (f) spatial contraction and expansion of a color system,
 and (g) resonance production.  They are interlaced by intricate connections
described in this paper. One may regard the approximation of
nonperturbative QCD  by color mutation as a gross simplification. We
conjecture that the final realistic result will be insensitive to the
details of how well many-body QCD is approximated, since the model is
constrained by so many other aspects of soft interaction.  This work
demonstrates the importance of how each and every one of those
separate components of the overall structure must work together in
order to achieve the scaling behavior of local fluctuations that is observed.

\begin{center}
\subsubsection*{Acknowledgment}
\end{center}

We are grateful to Prof.\ W.\ Kittel for supplying us with the NA22 data and
valuable answers to our questions.  This work was supported in part by U.S.
Department of Energy under Grant No. DE-FG03-96ER40972.

\vspace*{2cm}


\begin{figure}
\centerline{\epsfbox{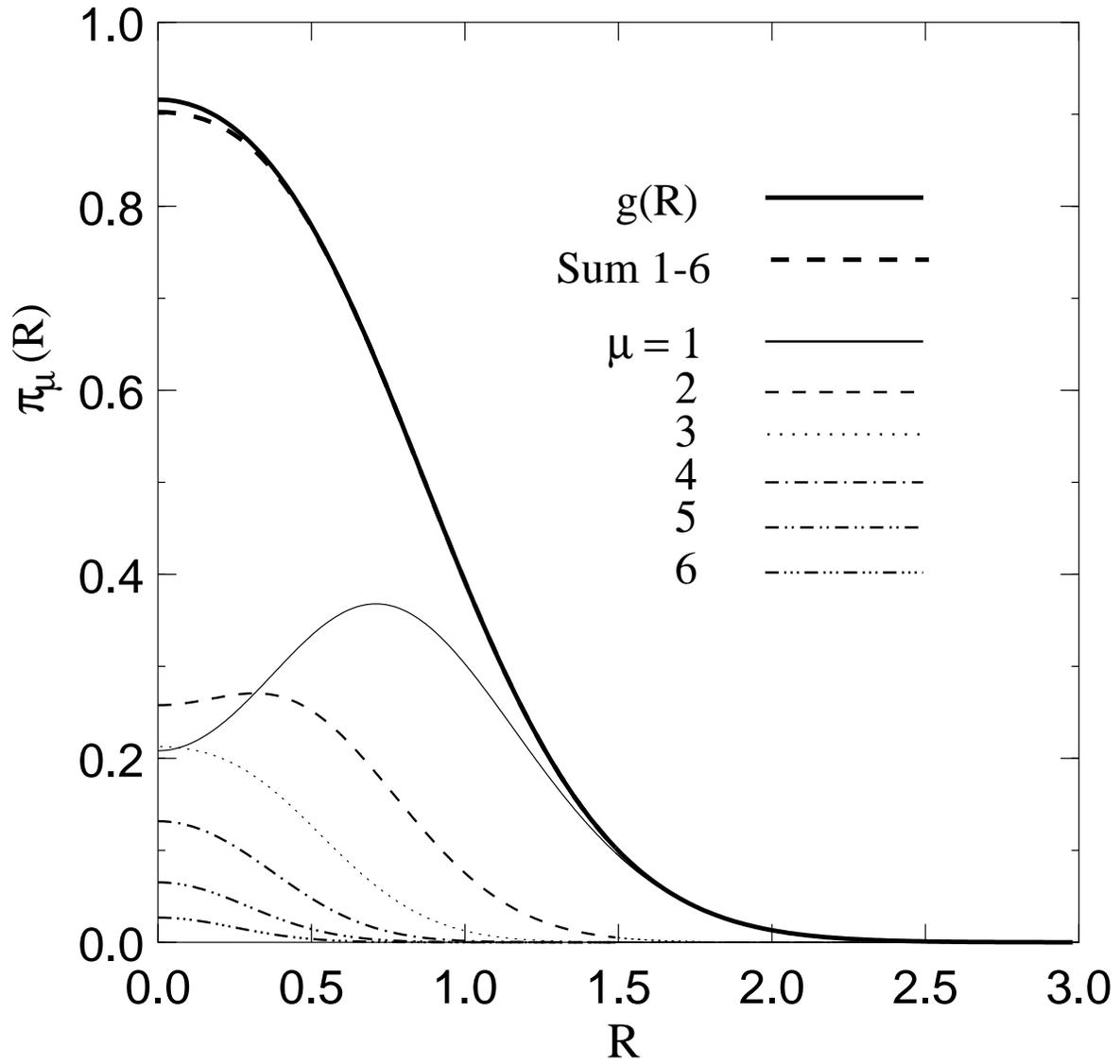}}
\label{fig:pimu_R}
\caption{Probability of having $\mu$-cut Pomerons at $R$.}
\end{figure}

\begin{figure}
\centerline{\epsfbox{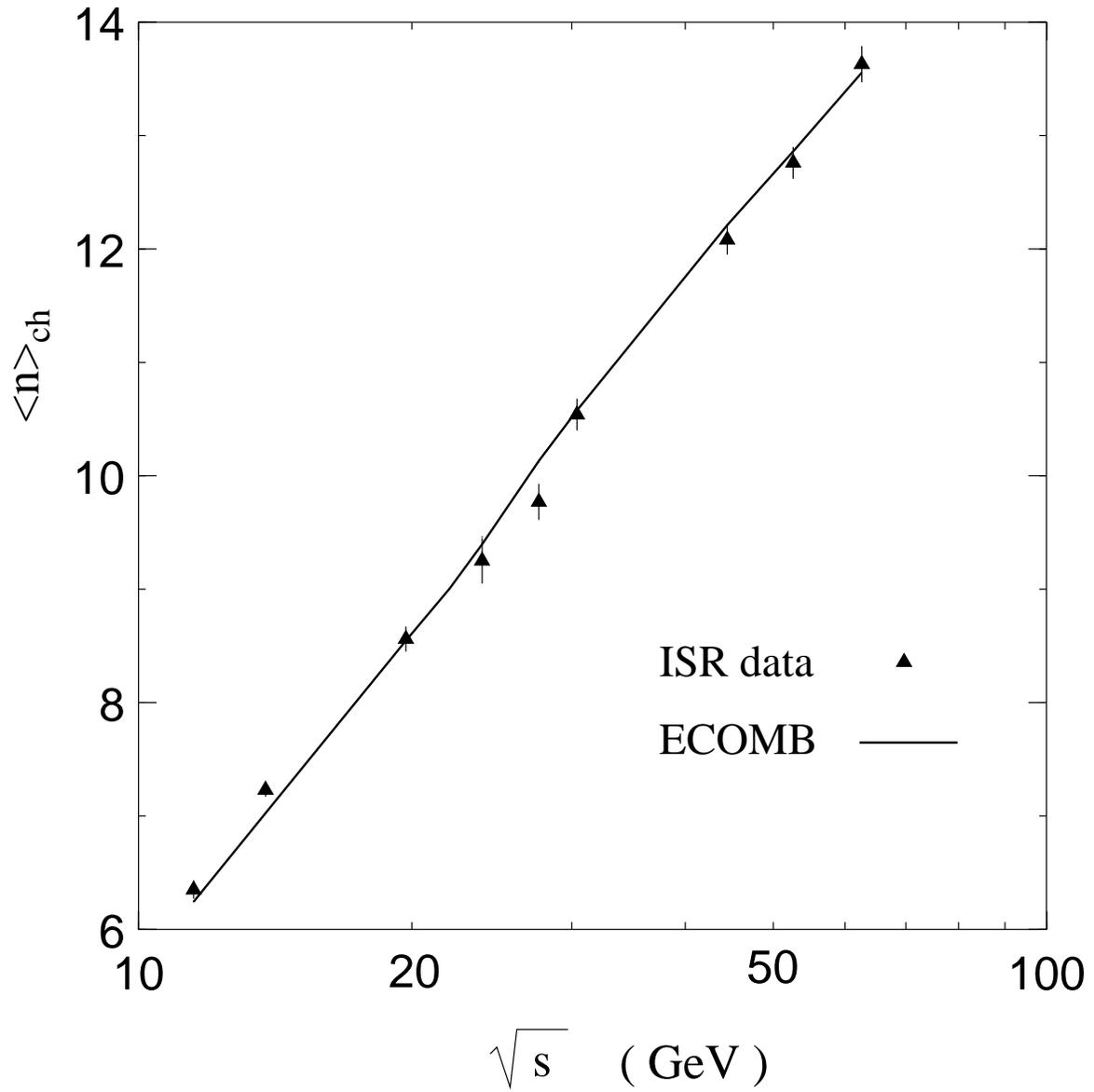}}
\label{fig:n_ch}
\caption{Average charge multiplicity as a function of cm energy. The data are from [19].}
\end{figure}

\begin{figure}
\centerline{\epsfbox{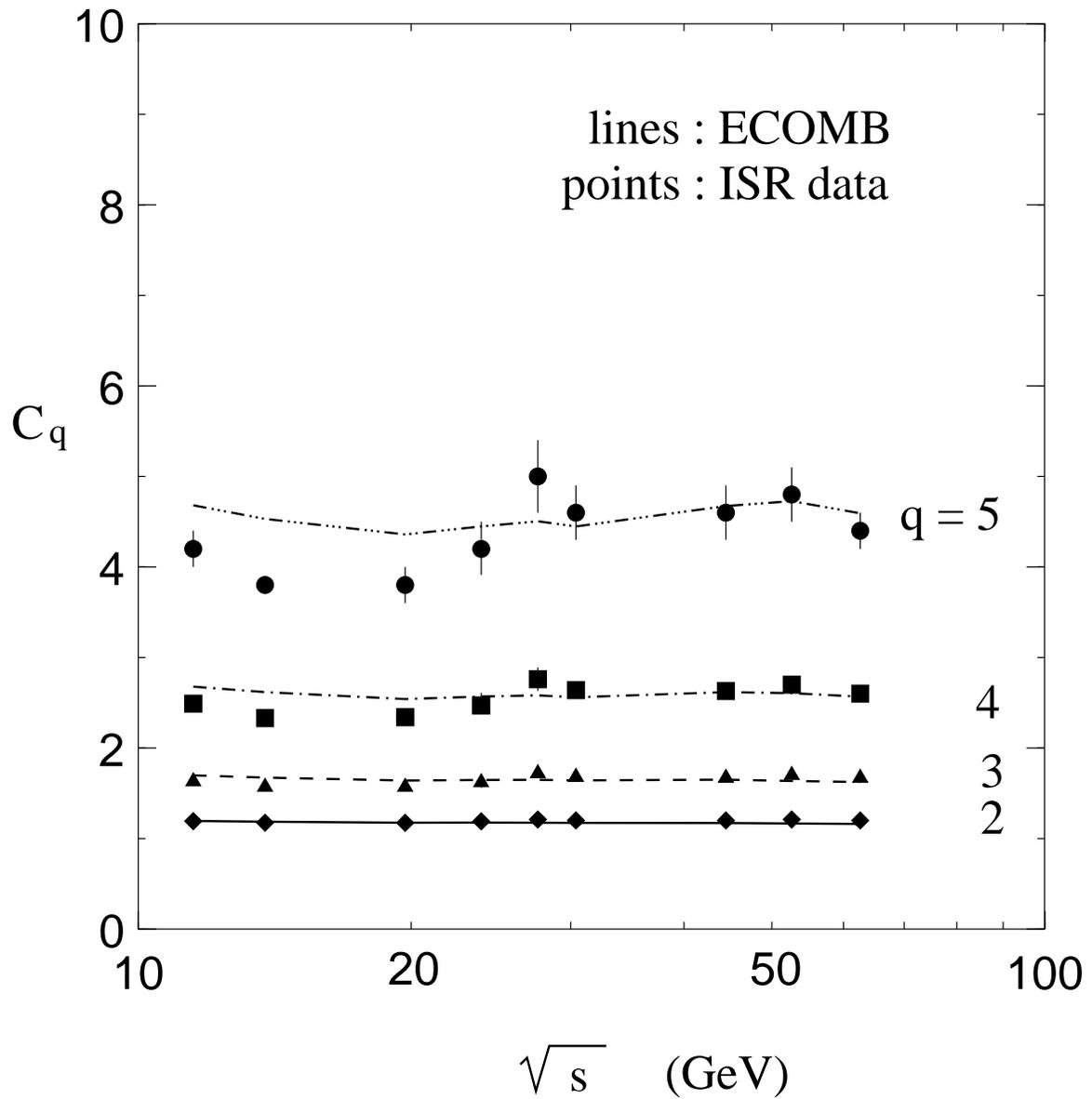}}
\label{fig:c_q}
\caption{Standard moments of the multiplicity distribution.  The data are from [19].}
\end{figure}

\begin{figure}
\centerline{\epsfbox{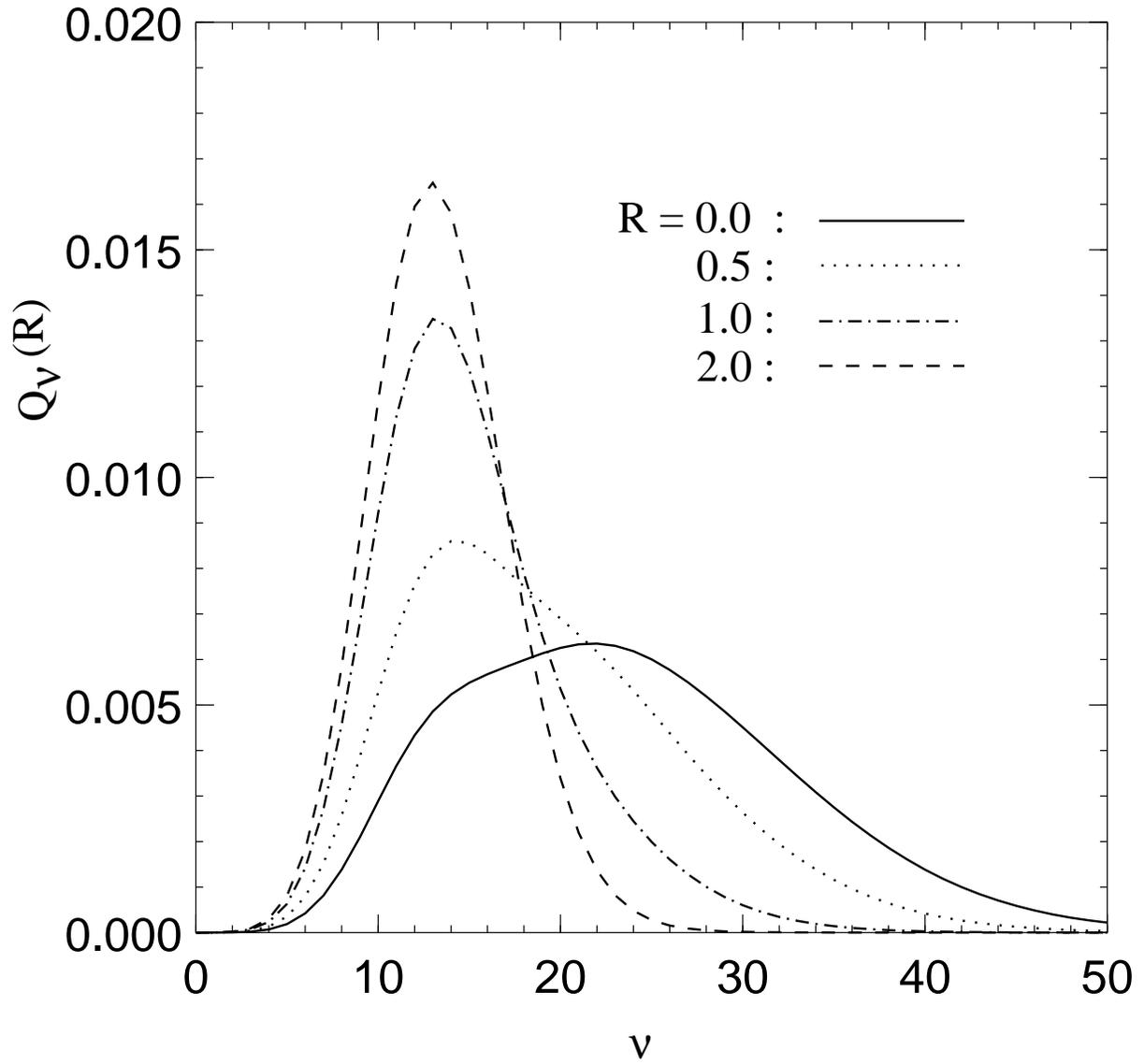}}
\label{fig:tgeom}
\caption{The distributions of parton numbers at various fixed $R$.}
\end{figure}

\begin{figure}
\centerline{\epsfbox{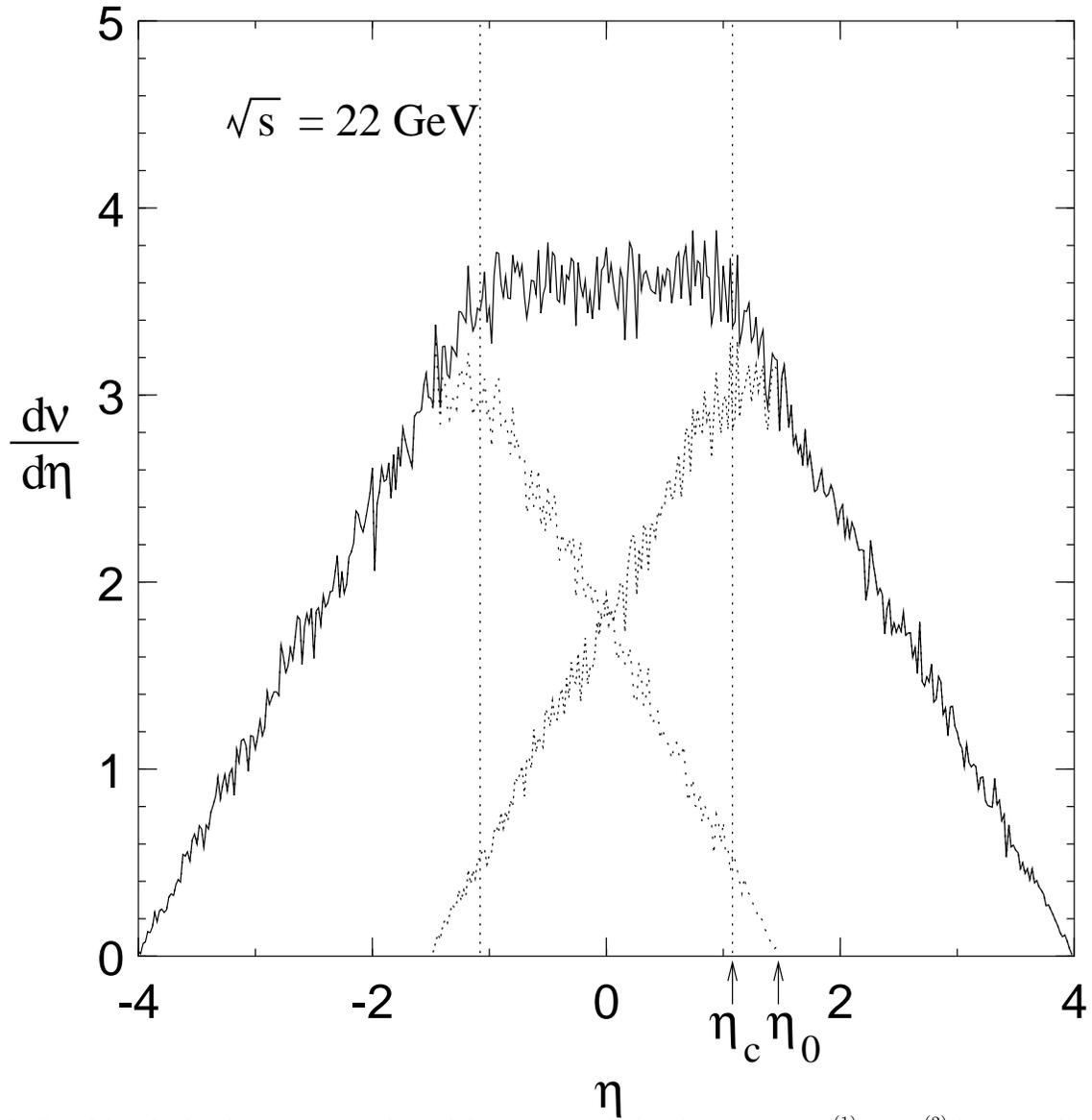}}
\label{fig:tyin}
\caption{Initial distributions of partons in rapidity $\eta$.
The solid line is the sum of $\rho^{(1)}$ and $\rho^{(2)}$ in dotted
lines.}
\end{figure}

\begin{figure}
\centerline{\epsfbox{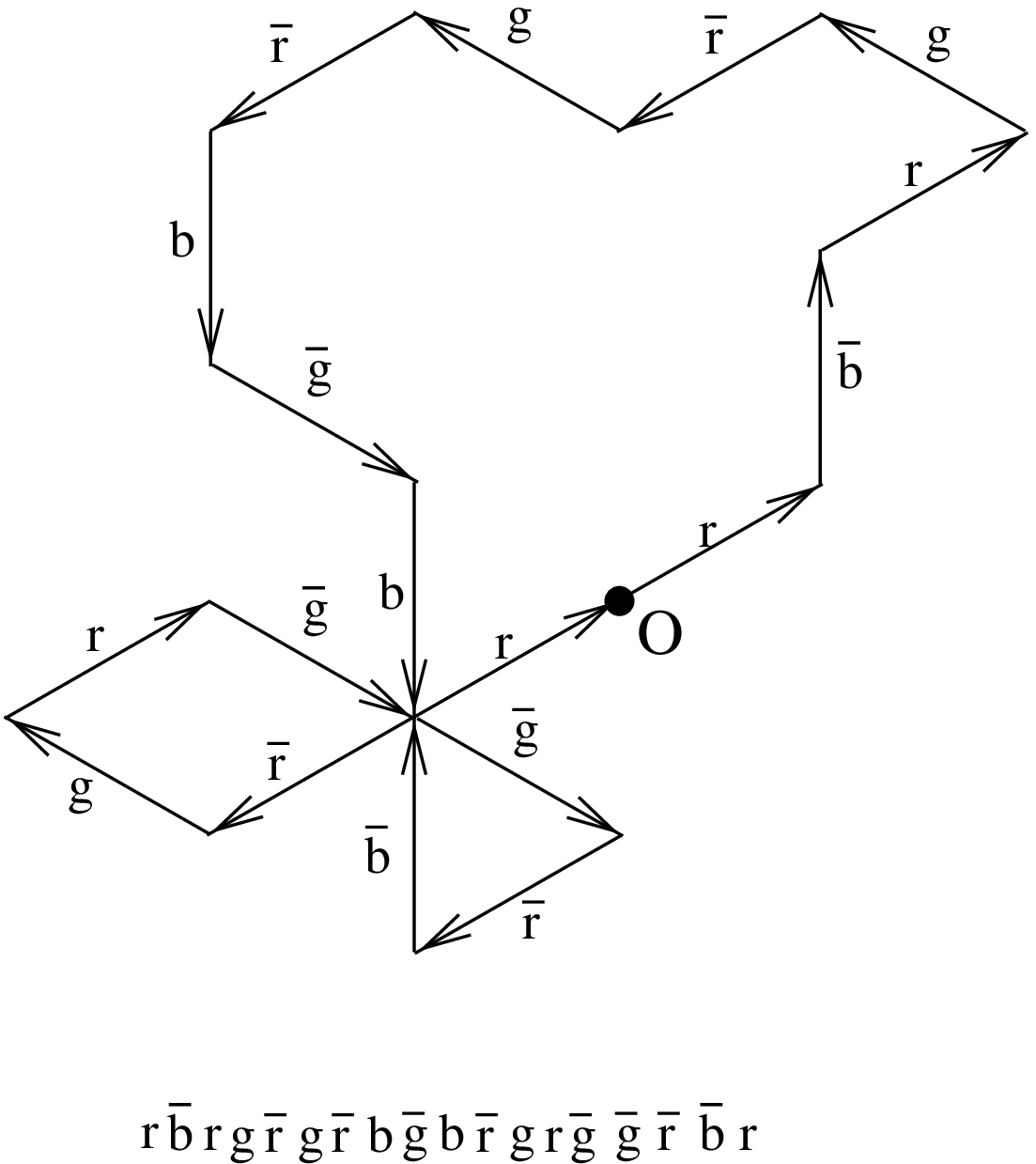}}
\label{fig:cpath}
\caption{A sample path in color space for a configuration
of partons arranged in the array shown.}
\end{figure}

\begin{figure}
\centerline{\epsfbox{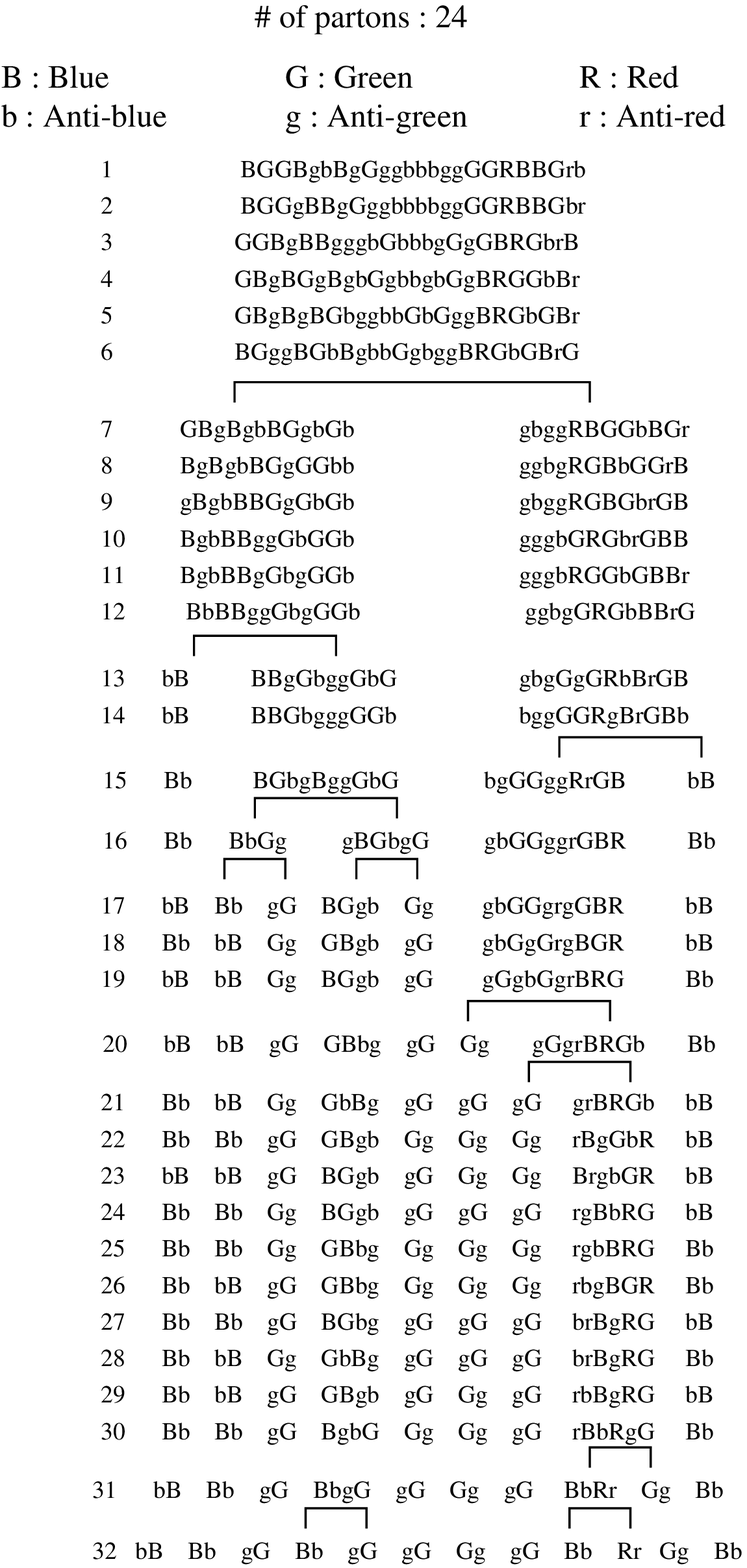}}
\label{fig:branch}
\caption{An example of a branching process that starts
with 24 partons. The numbers in the first column denote the number
of time steps taken for the configuration in any particular row.}
\end{figure}

\begin{figure}
\centerline{\epsfbox{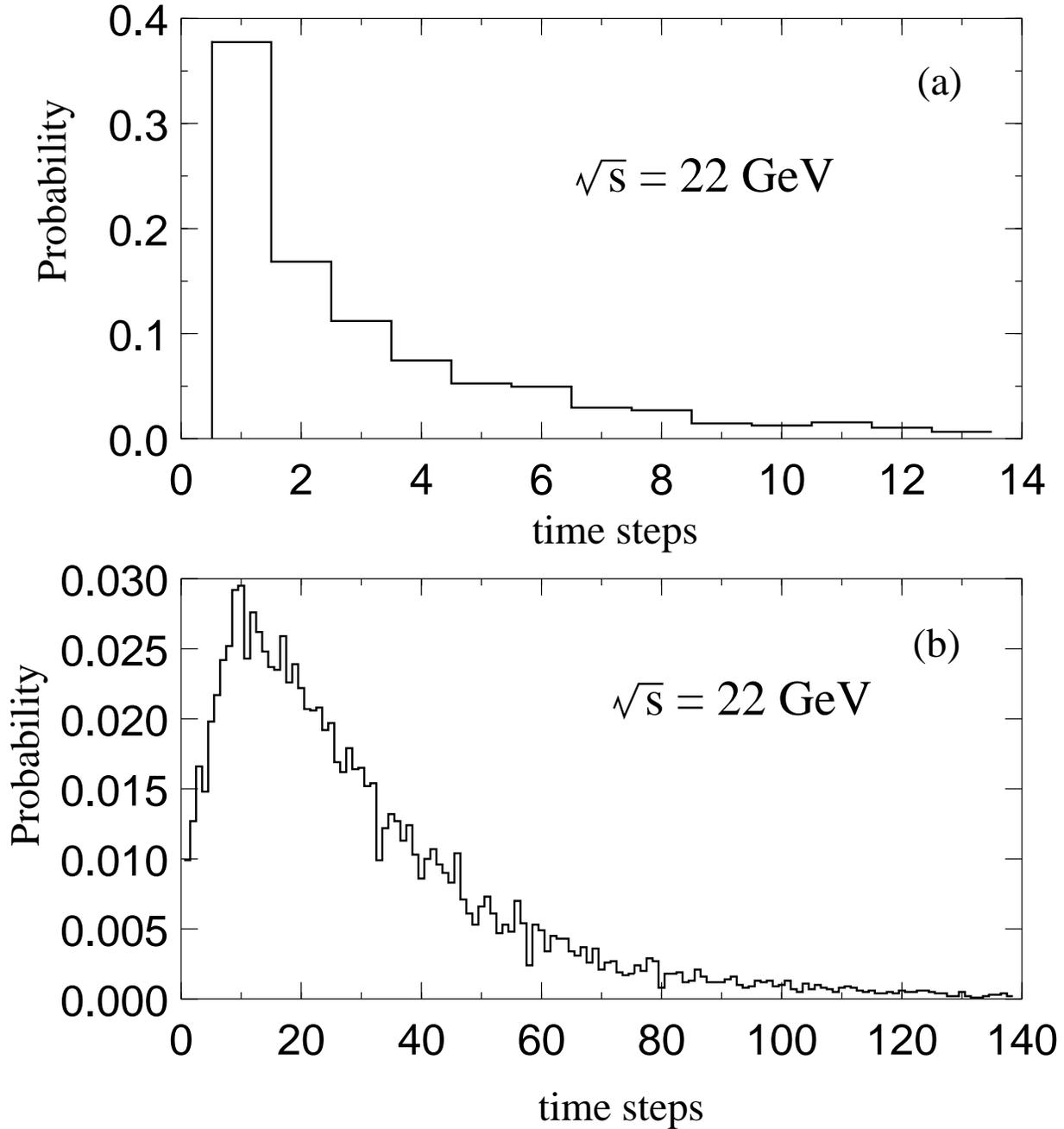}}
\label{fig:nevolut}
\caption{(a) The distribution of times steps taken for 20
initial partons before a fission process occurs. (b) The distribution of
time steps taken for the 20 partons to evolve to the end when
branching terminates.}
\end{figure}

\begin{figure}
\centerline{\epsfbox{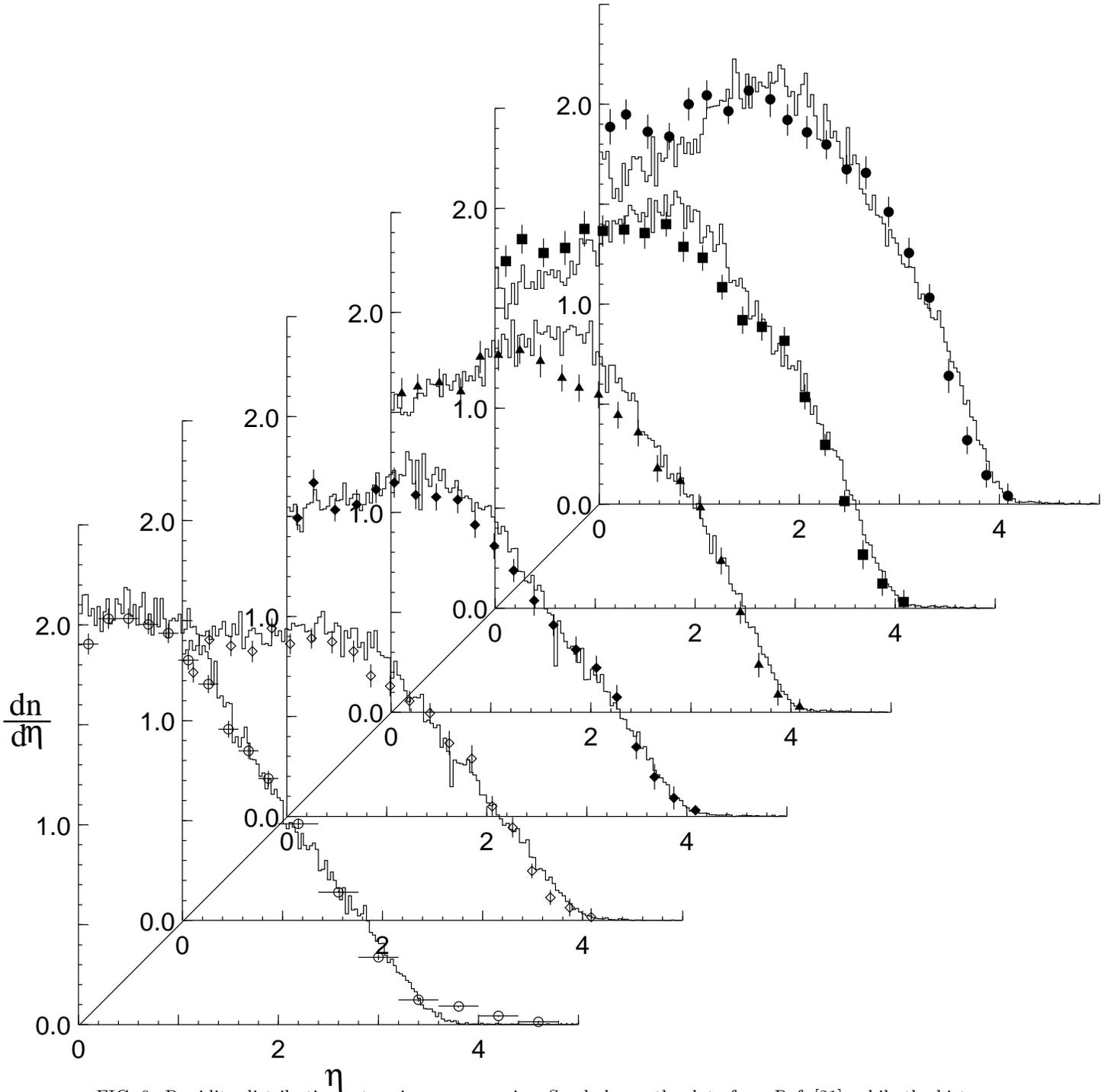}}
\label{fig:eta}
\caption{Rapidity distributions at various cm energies.
Symbols are the data from Ref.\ [21], while the histograms are the
result of ECOMB.}
\end{figure}

\begin{figure}
\centerline{\epsfbox{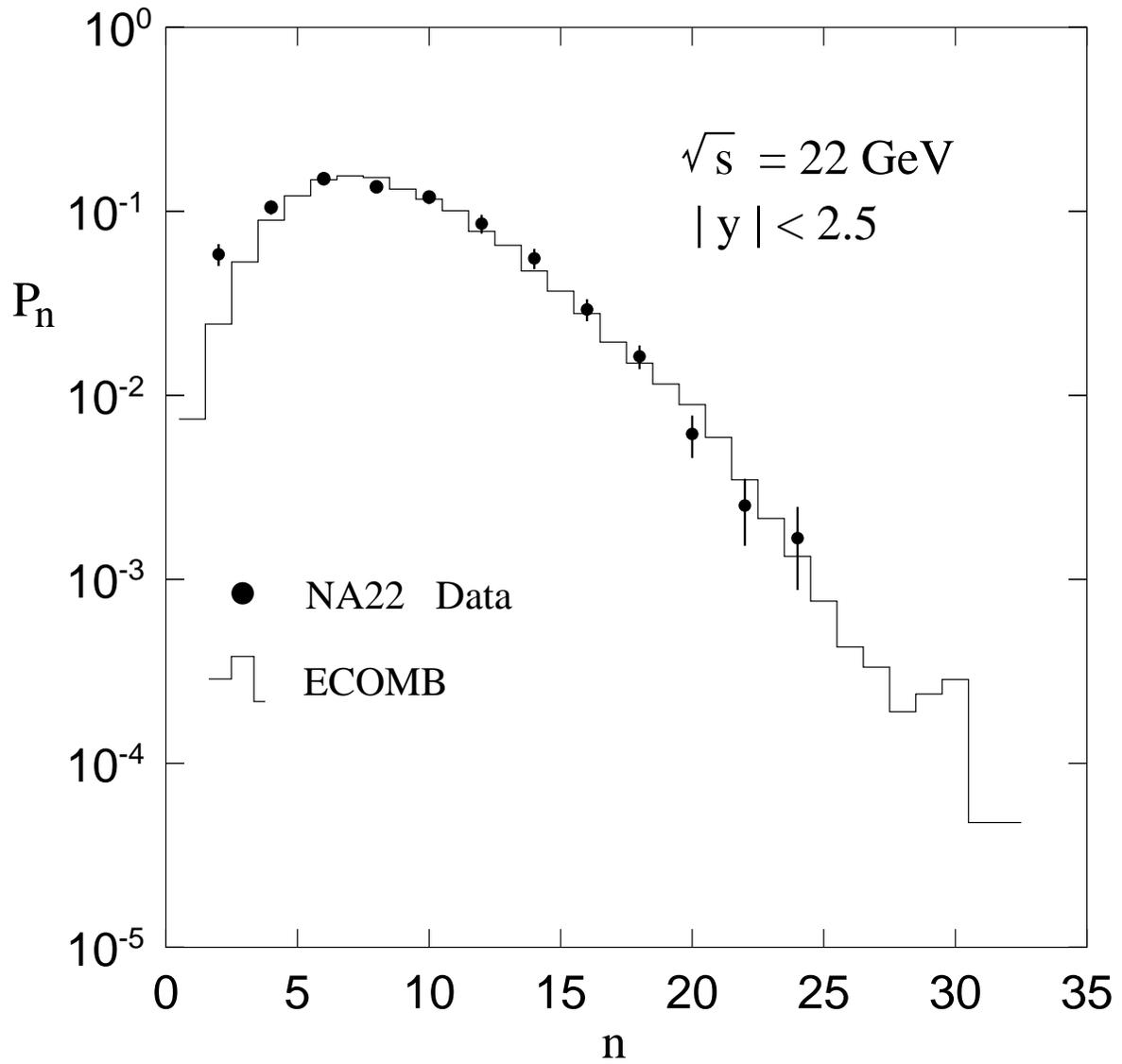}}
\label{fig:pn}
\caption{Multiplicity distributions of charged particles.
The data are from Ref.\ [22].}
\end{figure}

\begin{figure}
\centerline{\epsfbox{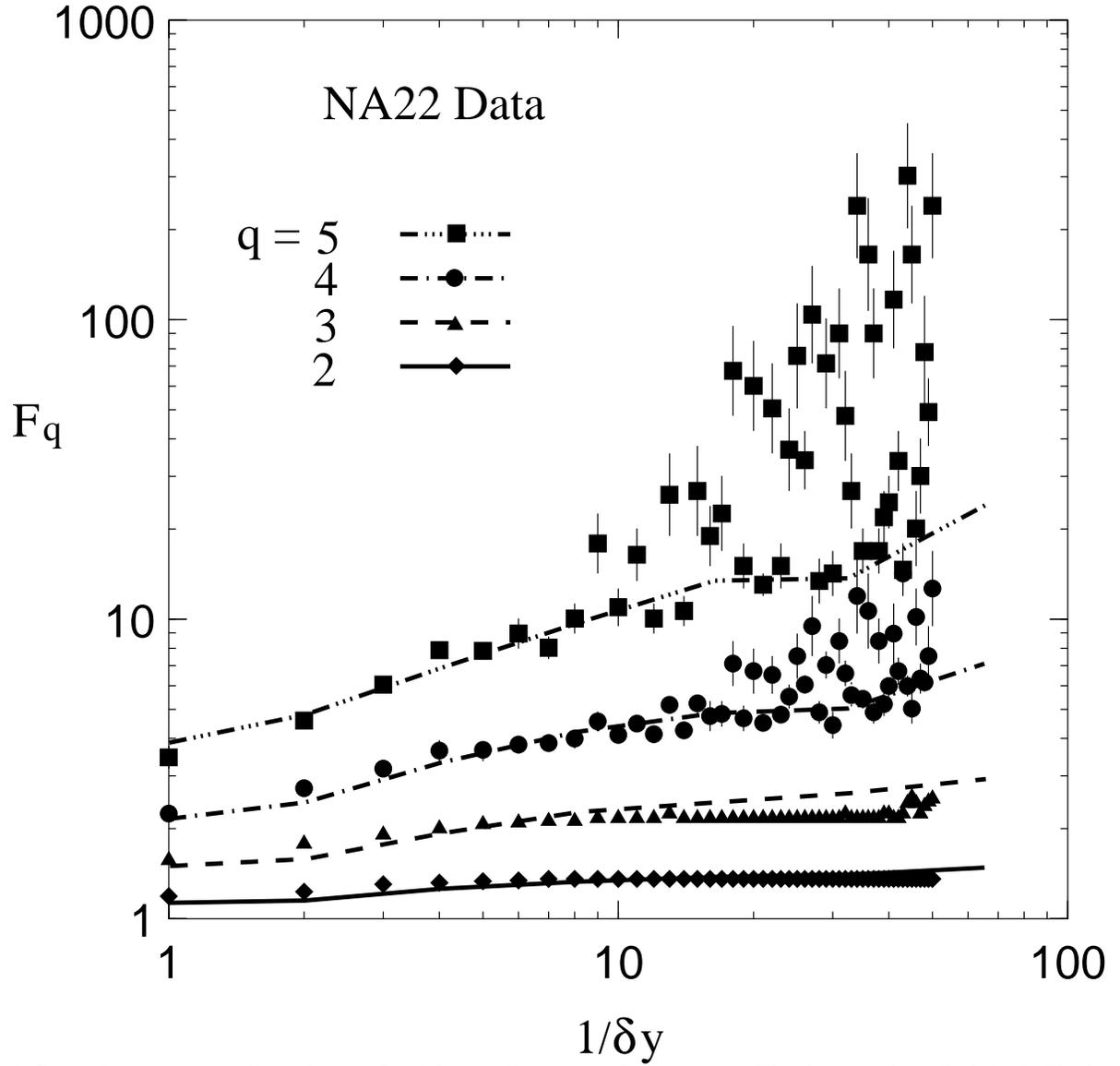}}
\label{fig:fq_m}
\caption{Intermittency data of normalized factorial
moments for $q=2-5$. The data are from Ref. [25]. The lines are
determined from ECOMB.}
\end{figure}

\end{document}